\documentclass[%
 reprint,
nofootinbib,
 amsmath,amssymb,
 aps,onecolumn
]{revtex4-1}

\usepackage{graphicx}
\usepackage{subcaption}
\usepackage{dcolumn}
\usepackage{bm}
\usepackage{hyperref}
\usepackage{cancel}
\usepackage{physics}
\usepackage{xcolor}

\begin{document}


\title{Quasitopological electromagnetism and black holes}

\author{Adolfo Cisterna}
 \affiliation{Vicerrector\'ia Acad\'emica, Toesca 1783, Universidad Central de Chile, Santiago, Chile and TIFPA - INFN, Via Sommarive 14, 38123 Povo (TN), Italy.}
 \email{adolfo.cisterna@ucentral.cl}
\author{Gaston Giribet}%
\affiliation{%
 Physics  Department,  University  of  Buenos  Aires  \&  IFIBA-CONICET, Ciudad  Universitaria,  pabell\'on  1,  1428,  Buenos  Aires,  Argentina.}
 \email{gaston@df.uba.ar}

\author{Julio Oliva}
\affiliation{
Departamento de F\'isica, Universidad de Concepci\'on, Casilla, 160-C, Concepci\'on, Chile.
}%
\email{juoliva@udec.cl}
\author{Konstantinos Pallikaris}
\affiliation{%
 Laboratory of Theoretical Physics, Institute of Physics, University of Tartu, W. Ostwaldi 1, 50411 Tartu, Estonia.
}%
\email{konstantinos.pallikaris@ut.ee}

\begin{abstract}
In this paper we extend the quasitopological electromagnetism, recently introduced by H.-S. Liu \textit{et al.} [arXiv:1907.10876], to arbitrary dimensions by introducing a fundamental $p$-form field. This allows us to construct new dyonic black hole solutions in odd dimensions, as well as regular $D$-dimensional black holes and solitons. The three-dimensional system consists of a Maxwell field interacting with a scalar field, leading to a deformation of the Ba\~nados-Teitelboim-Zanelli black hole. We present the general formulas defining the black hole solutions in arbitrary dimensions in Lovelock theory and explore the thermal properties of the asymptotically anti–de Sitter black holes in the gravitational framework of general relativity. In five dimensions, the latter black holes possess a rich phase space structure in the canonical ensemble, giving rise to as many as five different black hole phases at a fixed temperature, for a given range of the parameters.

\end{abstract}

\pacs{Valid PACS appear here}
\maketitle


\section{\label{sec:intro}Introduction}

One can think of the Einstein--Hilbert action as the higher--dimensional continuation of the two--dimensional Euler characteristic. This approach is actually useful when it comes to the problem of generalizing field theories. In fact, one can keep going in the same direction and define higher--curvature gravity theories by extending the $2k$--dimensional Chern-Weil topological invariants to $D$ dimensions, with $k =1, \, 2, \, 3, \, \ldots \leq \lfloor D/2\rfloor$, where $\lfloor\,\rfloor$ denotes the floor function. The theory obtained in this way is known as the Lovelock theory of gravity \cite{Lovelock}, and it is the best understood model involving higher--curvature couplings, usually employed to investigate the effects of higher-curvature terms in the context of AdS/CFT correspondence \cite{Brigante:2007nu}.

Recently, similar ideas have been explored in the case of Abelian gauge theories: In \cite{Liu:2019rib}, Liu \textit{et al.} introduced the notion of quasitopological electromagnetism, extending the Einstein--Maxwell theory by introducing new terms in the Lagrangian, which are related to topological invariants in a specific fashion. These new terms are built out of the Maxwell 2--form $F_{[2]}=dA_{[1]}$ and the metric tensor $g$. They involve polynomials of the form
\begin{equation}
V_{[2k]}=F_{[2]}\wedge F_{[2]}\wedge \ldots \wedge F_{[2]},
\end{equation}
with $k\leq \lfloor D/2\rfloor$ factors of the field strength 2--form $F_{[2]}$. We observe that polynomials $V_{[2k]}$ resemble the Pontryagin densities. In fact, in even dimensions $D=2k$, the integral of such a $D$--form is purely topological. In arbitrary dimensions, on the contrary, one may introduce these $2k$--forms in a way that does affect the dynamics of the classical theory. This can be achieved by considering the squared norm
\begin{equation}\label{La2}
U^{(k)}_{[D]}\sim |V_{[2k]}|^2\sim V_{[2k]}\wedge{}\ast\! V_{[2k]},
\end{equation}
with the case $k=1$ corresponding to the usual kinetic term of the Maxwell theory. In general, these invariants have nonvanishing contributions to the field equations. Such a theory has been dubbed quasitopological electromagnetism, and there are two reasons for such a name: first, the topological origin of its building blocks, the forms $V_{[2k]}$. Second, notice that for static and either purely electric or purely magnetic configurations, the spectrum of solutions coincides with the corresponding spectrum of standard Maxwell theory. Interesting phenomena emerge when dyons are considered, though.

In this paper, we consider a natural generalization of these quasitopological models by introducing, in addition to the Abelian gauge field $A_{[1]}$, a higher-rank $(p-1)$--form field $B_{[p-1]}$, whose field strength we will denote by $H_{[p]}=dB_{[p-1]}$. In specific cases, this new field can have different physical interpretations: for example, it might resemble the higher-rank fields appearing in string theory, such as the ubiquitous Kalb--Ramond 2-form field $B_{[2]}$, the Ramond--Ramond $p$-forms of the Type II theories, or the 3--form field of 11-dimensional supergravity. An interesting \emph{Ansatz} for this field is to consider it as purely magnetic, wrapped around the horizon geometry of a static black brane solution. As shown below, having couplings between the Maxwell field and $p$-forms allows for more general configurations than those considered in \cite{Liu:2019rib} for the single field model. For example, as we will see, the presence of $p$-forms permits odd-dimensional versions of the model in which dyonic black holes can also be analytically studied.

This paper is organized as follows: In Sec. II, we introduce the generalized quasitopological theory and extract the field equations. In Sec. III, we derive static, dyonic black hole solutions of the theory coupled to higher-curvature Lovelock gravity. We analyze the geometrical and the thermodynamical properties of the solutions, focusing our attention on asymptotically anti--de Sitter (AdS) black holes, which exhibit a very rich variety of configurations. We also study the horizon structure of the four--dimensional case, as well as the possibility of obtaining nonsingular dyonic solutions. We also pay special attention to the case $D=3$, where the $B_{[0]}$ field appears as a backreacting scalar, leading to a deformation of the Ba\~nados-Teitelboim-Zanelli (BTZ) spacetime.

\section{Quasitopological field theory\label{sec:II}}

Extending the idea of \cite{Liu:2019rib}, we can construct similar structures as the one in \eqref{La2} by using the field strength $H_{[p]}$. More precisely, we can consider the quantities
\begin{eqnarray}
\mathcal F_{[2k]}&=&F_{[2]}\wedge F_{[2]}\wedge\ldots \wedge F_{[2]}\ ,\qquad \qquad \ k\leq \lfloor D/2\rfloor\ ,\nonumber\\
\mathcal H_{[pk]}&=&H_{[p]}\wedge H_{[p]}\wedge\ldots \wedge H_{[p]}\ ,\qquad k\leq \lfloor D/p\rfloor\ ,\nonumber\\
\mathcal{FH}_{[2k+p\ell]}&=&\mathcal F_{[2k]}\wedge \mathcal H_{[p\ell]}\ ,\qquad \qquad \ \  2k+p\ell\leq D\ .
\end{eqnarray}
With these at hand, we can introduce squared norms using the Hodge product; namely $|\mathcal{F}_{[2k]}|^2$, $|\mathcal{H}_{[pk]}|^2$, and $|\mathcal{FH}_{[2k+p\ell]}|^2$. In component notation, these read 
\begin{eqnarray}
|\mathcal{F}_{[2k]}|^2&\sim &\delta^{\alpha_1\ldots \alpha_{2k}}_{\beta_1\ldots \beta_{2k}}F_{\alpha_1 \alpha_2}F_{\alpha_3\alpha_4}\ldots F_{\alpha_{2k-1}\alpha_{2k}}F^{\beta_1 \beta_2}F^{\beta_3\beta_4}\ldots F^{\beta_{2k-1}\beta_{2k}}\ ,\nonumber\\
|\mathcal{H}_{[pk]}|^2&\sim&\delta^{\alpha_1\ldots \alpha_{pk}}_{\beta_1\ldots \beta_{pk}}H_{\alpha_1\ldots \alpha_p}\ldots H_{\ldots \alpha_{pk}}H^{\beta_1\ldots \beta_p}\ldots H^{\ldots \beta_{pk}}\ ,\nonumber\\
|\mathcal{FH}_{[2k+p\ell]}|^2&\sim&\delta^{\alpha_1\ldots \alpha_{2k+p\ell}}_{\beta_1\ldots \beta_{2k+p\ell}}F_{\alpha_1\alpha_2}H_{\alpha_3\ldots \alpha_{p+2}}\ldots F_{\ldots}H_{\ldots \alpha_{2k+p\ell}}F^{\beta_1\beta_2}H^{\beta_3\ldots \beta_{p+2}}\ldots F^{\ldots}H^{\ldots \beta_{2k+p\ell}}\ ,
\end{eqnarray}
where $\delta^{\alpha_1\ldots \alpha_{2k}}_{\beta_1\ldots \beta_{2k}}$ stands for the rank-$4k$ skew-symmetric Kronecker delta. There are, of course, other possibilities in addition to these squared norms. For example, one can consider terms of the form $\mathcal{F}_{[2k]}\wedge\ast \mathcal{H}_{[p\ell]}$ with $2k=p\ell$ and $k\leq \lfloor D/2\rfloor$, $\mathcal{F}_{[2k]}\wedge \ast \mathcal{FH}_{[2q+p\ell]}$ with $p\ell=2(k-q)$ and $k\leq \lfloor D/2\rfloor$, as well as $\mathcal{H}_{[pk]}\wedge \ast \mathcal{FH}_{[2q+p\ell]}$ with $2q=p(k-\ell)$ and $k\leq\lfloor D/p\rfloor$. In general, all of these invariants would contribute to the field equations and, therefore, \emph{a priori} they should be included in the action. 

However, we will be interested in configurations of the form 
\begin{equation}
F_{\mu\nu}\sim a'(r)\delta^{x^0x^1}_{\mu\nu}\ ,\qquad H_{\alpha_1\ldots \alpha_p}\sim\delta^{x^{2}\ldots x^{D}}_{\alpha_1\ldots \alpha_p}\ ,
\end{equation}
with $p=D-2$, in which $F_{\mu\nu}$ is purely electric and $H_{\alpha_1\ldots \alpha_p}$ purely magnetic. One can show that the only nonvanishing terms for such configurations would be the kinetic terms $|\mathcal{F}_{[2]}|^2\sim F_{\mu\nu}F^{\mu\nu}$ and $|\mathcal{H}_{[p]}|^2$, and the interacting term $|\mathcal{FH}_{[D]}|^2$ written above. For that reason, it will be sufficient for us to consider a $D$--dimensional action of the form
\begin{equation}
I_D\!\left[g_{\mu\nu},A_\mu,B_{\alpha_1\ldots \alpha_{p-1}}\right]=\int \,d^Dx\,\sqrt{-g}\,\mathcal{L}_{\mathrm{Lov}}-\int \,d^Dx\,\sqrt{-g}\, \left[\dfrac{1}{4}F^2+\dfrac{1}{2p!}H^2+\alpha \mathcal{L}_{\mathrm{int}}\right]\ ;\label{eq:action}
\end{equation}
where $F^2=F_{\mu\nu}F^{\mu\nu}$ and $H^2=H_{\alpha_1\ldots \alpha_p}H^{\alpha_1\ldots \alpha_p}$, with the interaction term given by
\begin{equation}
\mathcal{L}_{\mathrm{int}}=\delta^{\alpha_1\ldots \alpha_D}_{\beta_1\ldots \beta_D}F_{\alpha_1\alpha_2}H_{\alpha_3\ldots \alpha_D}F^{\beta_1\beta_2}H^{\beta_3\ldots \beta_D}\ .
\end{equation}
Here the coupling constant $\alpha$ has mass dimension $-2$. As is well-known, the Lovelock Lagrangian reads 
\begin{equation}
\mathcal{L}_{\mathrm{Lov}}=\sum_{k=0}^{\lfloor D/2\rfloor}\dfrac{a_k}{2^{k}}\delta^{\beta_1\ldots \beta_{2k}}_{\alpha_1\ldots \alpha_{2k}}R^{\alpha_1\alpha_2}{}_{\beta_1\beta_2}\ldots R^{\alpha_{2k-1}\alpha_{2k}}{}_{\beta_{2k-1}\beta_{2k}}\ , \label{LovLag}
\end{equation}
where the coefficients $a_k$ are dimensionful coupling constants of mass dimensions $D-2k$.

The field equations coming from the action principle (\ref{eq:action}) read
\begin{eqnarray}
\mathcal{G}_{\mu\nu}&=&\sum_{k=0}^{\lfloor D/2 \rfloor}{a_k}\mathcal E_{\mu\nu}^{(k)}-\dfrac{1}{2}F_{\mu\rho}F_{\nu}{}^\rho+\dfrac{1}{8}g_{\mu\nu}F^2-\dfrac{1}{4}\mathcal{B}_{\mu\nu}-\dfrac{\alpha}{2}g_{\mu\nu}\mathcal{L}_{\mathrm{int}}=0\ ,\label{eq:eomg}\\
\mathcal{M}^{\mu}&=&\nabla_{\nu}F^{\nu\mu}-4\alpha \delta^{\mu\nu \alpha_1\ldots \alpha_p}_{\beta_1\ldots \beta_D}H_{\alpha_1\ldots \alpha_p}\nabla_{\nu}(F^{\beta_1\beta_2}H^{\beta_3\ldots \beta_D})=0\ ,\label{eq:eomAfield}\\
\mathcal{K}^{\alpha_1\ldots \alpha_{p-1}}&=&\nabla_{\mu}H^{\mu\alpha_1\ldots \alpha_{p-1}}+2\alpha p!\delta^{\mu\nu\rho \alpha_1\ldots \alpha_{p-1}}_{\beta_1\ldots \beta_D}F_{\mu\nu}\nabla_{\rho}(F^{\beta_1\beta_2}H^{\beta_3\ldots \beta_D})=0\ ,\label{eq:eomHfield}
\end{eqnarray}
The Lovelock tensors $\mathcal E_{\mu\nu}^{(k)}$ are defined as
\begin{equation}
\mathcal E_{\mu\nu}^{(k)}=-\dfrac{1}{2^{k+1}}\delta^{\rho\alpha_1\ldots \alpha_{2k}}_{\beta_1\ldots \beta_{2k}(\mu}g_{\nu)\rho}R^{\beta_1\beta_2}{}_{\alpha_1\alpha_2}\ldots R^{\beta_{2k-1}\beta_{2k}}{}_{\alpha_{2k-1}\alpha_{2k}}\ ,
\end{equation}
while the energy--momentum tensor for $B_{[p-1]}$ reads
\begin{equation}
\mathcal{B}_{\mu\nu}=\dfrac{1}{(p-1)!}H_{\mu\alpha_1\ldots \alpha_{p-1}}H_{\nu}{}^{\alpha_1\ldots \alpha_{p-1}}-\dfrac{1}{(p!)^2}\delta^{\alpha_1\ldots \alpha_p\rho}_{\beta_1\ldots \beta_p(\mu}g_{\nu)\rho}H_{\alpha_1\ldots \alpha_p}H^{\beta_1\ldots \beta_p} \ .
\end{equation}

An interesting comment is now in order. For the contribution of the interaction part of the Lagrangian, $\mathcal{L}_{\mathrm{int}}$, to the field equations one would have expected a term of the form
\begin{equation}
\dfrac{1}{\sqrt{-g}}\dfrac{\delta(\sqrt{-g}\mathcal{L}_{\mathrm{int}})}{\delta g^{\mu\nu}}=X_{\mu\nu}-\dfrac{1}{2}g_{\mu\nu}\mathcal{L}_{\mathrm{int}}\ .
\end{equation}
Nevertheless these Lagrangians fulfill the identity
\begin{equation}
0\equiv \delta^{\alpha_1\ldots \alpha_D}_{\beta_1\ldots \beta_D}F_{[\alpha_1\alpha_2}H_{\alpha_3\ldots \alpha_D}F^{\beta_1\beta_2}H^{\beta_3\ldots \beta_D}g_{\mu]\nu}=-X_{\mu\nu}+g_{\mu\nu}\mathcal{L}_{\mathrm{int}}\ , 
\end{equation}
and therefore,
\begin{equation}
\dfrac{1}{\sqrt{-g}}\dfrac{\delta(\sqrt{-g}\mathcal{L}_{\mathrm{int}})}{\delta g^{\mu\nu}}=\dfrac{1}{2}g_{\mu\nu}\mathcal{L}_{\mathrm{int}}\ ,   \label{iden}
\end{equation}
which allows us to recast the energy--momentum tensor of the interaction term in a simpler form, leading to the field equations (\ref{eq:eomg})--(\ref{eq:eomHfield}). In what follows, we study static, spherically symmetric, dyonic black hole solutions to these field equations.

\section{Exact solutions\label{sec:III}}

\subsection{\boldmath $D$--dimensional solutions} 

Let us now construct exact solutions to the theory defined by (\ref{eq:action}), which despite its apparent complexity can be integrated explicitly even when both electric and magnetic charges are present. Consider the static spherically symmetric metric
\begin{equation}
ds^2=-G(r)dt^2+\dfrac{dr^2}{G(r)}+r^2d\Sigma^{2}_{D-2,\gamma} \ .\label{eq:metric}
\end{equation}
Here $d\Sigma_{D-2,\gamma}$ is the line element of a Euclidean manifold of constant curvature $\gamma=\pm1,0$. It will be useful to think about a local chart $\{x^{i}\}$, with $i=1,\ldots,p$, which leads to an intrinsic metric $\sigma_{ij}$ on the manifold $\Sigma_{D-2,\gamma}$, with determinant $\sigma$. This $(D-2)$--dimensional hypersurface will be dressed with a magnetic field proportional to its intrinsic volume form, $H_{[D-2]}\sim \mathrm{Vol}(\Sigma)$, namely
\begin{equation}
H_{\alpha_1\ldots \alpha_p}=q_m\sqrt{\sigma}\delta^{x^1\ldots x^p}_{\alpha_1\ldots \alpha_p}\label{eq:Btens}\ . 
\end{equation}
The Maxwell field will be purely electric,
\begin{equation}
F_{\mu\nu}=a'(r)\delta^{tr}_{\mu\nu}\ ,\label{eq:Ftens}
\end{equation}
where the prime stands for the derivative with respect to $r$.

In this \emph{Ansatz}, the Maxwell equations reduce to
\begin{equation}
r^{2p}[p a'(r)+r a''(r)]-8\alpha(p!)^2{q_m^2}[p{a'(r)}-ra''(r)]=0\ ,\qquad p=D-2\ ,\label{eq:maxwell}
\end{equation}
which has the solution 
\begin{equation}
a'(r)=\dfrac{q_e r^p}{r^{2p}+8\alpha(p!)^2q_m^2}\ .\label{eq:maxsolution}
\end{equation}
This equation demonstrates the screening of the electric field produced by the interaction with the magnetic component. The equations for the field $B_{[p-1]}$, on the other hand, are identically fulfilled in this \emph{Ansatz}. 

Therefore, it remains to solve the gravitational field equations: For a generic Lovelock theory, the field equations can be integrated in terms of a Wheeler-type polynomial \cite{Wheeler:1985qd}, which comes from the trivial integration of the first-order ordinary differential equation
\begin{equation}\label{tointegrate}
\dfrac{D-2}{2r^{D-2}}G(r)\partial_r\left(r^{D-1}\sum_{k=0}^{\lfloor D/2\rfloor }\tilde{a}_k\left(\frac{\gamma-G(r)}{r^2}\right)^{k}\right)=T_{tt}\ ,
\end{equation}
where
\begin{equation}
T^{t}_t=-\dfrac{1}{4}\left(\dfrac{q_m^2}{r^{2(D-2)}}+\dfrac{q_e^2}{r^{2(D-2)}+8\alpha q_m^2  \Gamma[D-1]^2}\right)\ .\label{eq:density}
\end{equation}
Since $\alpha$ is taken to be positive, $T_{tt}$ turns out to always be positive. For convenience, above we have introduced the rescaled coupling constants
\begin{equation}
\tilde{a}_0=\dfrac{a_0}{(D-1)(D-2)}\ ,\qquad \tilde{a}_1=a_1\ ,\qquad \tilde{a}_k=a_k\prod_{i=3}^{2k}(D-i)\ ,
\end{equation}
the last for $k>1$. It is worth mentioning that the upper limit of the sum in Eq. \eqref{tointegrate} can be extended to values higher than $\lfloor D/2 \rfloor$ in the context of  quasitopological gravity. Such models were originally introduced in the cubic case \citep{OlivaRay} (see also \citep{Myers-Robinson}), and were later extended to the quartic and quintic cases in \citep{Dehghani-Mann,Quintic} (see also the recent \citep{PablosRecursive}). These theories lead to second-order field equations in spherically symmetric spacetimes, with the same structure as those of Lovelock theories.

In order to present an explicit form of the solution that will permit us to study the main features introduced by the quasitopological Abelian fields, hereafter we restrict to Einstein gravity with a cosmological constant in arbitrary dimensions $D$; namely we fix the coupling constants as $a_k=\delta_k^1-2\Lambda\delta_k^0$ in the Lagrangian \eqref{LovLag}. This corresponds to setting $16\pi G\!_N=1$ in the usual normalization of the Einstein-Hilbert action with a bare cosmological constant $\Lambda$. In this case the $\mathcal{G}_{tt}=0$ component of the gravitational field equations reads
\begin{equation}
\dfrac{4\mathcal{G}_{tt}}{G(r)}=\dfrac{2(D-2)(D-3)\gamma}{r^2}-4\Lambda-\dfrac{q_m^2}{r^{2(D-2)}}-\dfrac{q_e^2}{r^{2(D-2)}+8\alpha\Gamma[D-1]^2q_m^2}-2(D-2)\dfrac{rG'(r)+(D-3)G(r)}{r^2} \label{eq:metfieldeval}\ ,
\end{equation}
leading to
\begin{equation}\label{EinsteinsGfunction2}
G(r)=\gamma-\dfrac{M}{(D-2)\sigma_\gamma r^{D-3}}-\dfrac{2\Lambda r^2}{(D-1)(D-2)}+\dfrac{q_m^2+ q_e^2\ {}_2F_1\bqty{1,\frac{D-3}{2(D-2)},\frac{3D-7}{2(D-2)},-\frac{8\alpha q_m^2\Gamma[D-1]^2}{r^{2(D-2)}}}}{2(D-2)(D-3)r^{2(D-3)}}\ .
\end{equation}
Here $_2F_1$ denotes Euler's hypergometric function. The integration constant $M$ is the Arnowitt-Deser-Misner (ADM) mass, and $\sigma_\gamma$ is the volume of the manifold $\Sigma_{(D-2),\gamma}$, which is equal to $\sigma_{1}=2\pi^{(D-1)/2}/\Gamma[(D-1)/2]$ for a hyperspherical horizon. The presence of hypergeometric functions in the black hole is reminiscent of what happens in Lovelock--Born--Infeld theory; see \cite{Ferraro} and references therein and thereof.

It is observed that, for a certain range of parameters, solution (\ref{EinsteinsGfunction2}) has positive roots, which can be multiple. These roots define the location of the Killing horizons. Besides, the metric is regular for all values of $r$ larger than the smallest positive root. This means that, for a certain set of parameters and coupling constants, the solution describes a static charged black hole. In the case of coincident roots, the near horizon geometry becomes AdS$_2 \times \sigma_\gamma$, while, as usual, the standard Rindler structure appears near the nondegenerate horizons. In the latter case, the black hole has nonvanishing Hawking temperature and nontrivial entropy. Below we explore the thermodynamics of the dyonic black holes in arbitrary dimensions. 

\subsection{Black hole thermodynamics}

For concreteness, let us focus on the asymptotically AdS solutions in general relativity. In this case, the Hawking temperature reads
\begin{equation}
    T=\frac{G'(r_+)}{4\pi}=\frac{r_+}{8(D-2)\pi}\left(\frac{2(D-2)(D-3)\gamma}{r_+^2}-4\Lambda-\frac{q_m^2}{r_+^{2(D-2)}}-\frac{q_e^2}{r_+^{2(D-2)}+8\alpha q_m^2\Gamma[D-1]^2}\right)\ ,\label{temperature}
\end{equation}
where $r=r_+$ is the location of the event horizon, defined as the largest root of the equation $G(r_+)=0$. The radius $r_+$ is, of course, a function of the integration constants $M,\ q_e,$ and $q_m$, as well as of the coupling constants $\Lambda$ and $\alpha $. The asymptotic behavior of the solution \eqref{EinsteinsGfunction2} shows that the matter distribution can be thought of as that of a localized object in AdS space, since
\begin{align}
    G(r)= &-\frac{2 \Lambda  r^2}{(D-2) (D-1)}+\gamma-\frac{M}{(D-2)\sigma_\gamma r^{D-3}}+\frac{q_e^2+q_m^2}{2 (D-3) (D-2)r^{2(D-3)}}\nonumber\\
    &-\frac{4 \alpha  q_e^2 q_m^2 \Gamma[D-1]^2}{(D-2) (3D-7)r^{2(2D-5)}}+\mathcal{O}\left(\frac{1}{r^{2(3D-7)}}\right)\ 
\end{align}
obeys the Brown-Teitelboim asymptotically AdS$_{D>3}$ boundary conditions. 
\begin{figure}[ht] 
  \begin{subfigure}[b]{0.5\textwidth}
    \centering
    \includegraphics[width=0.75\textwidth]{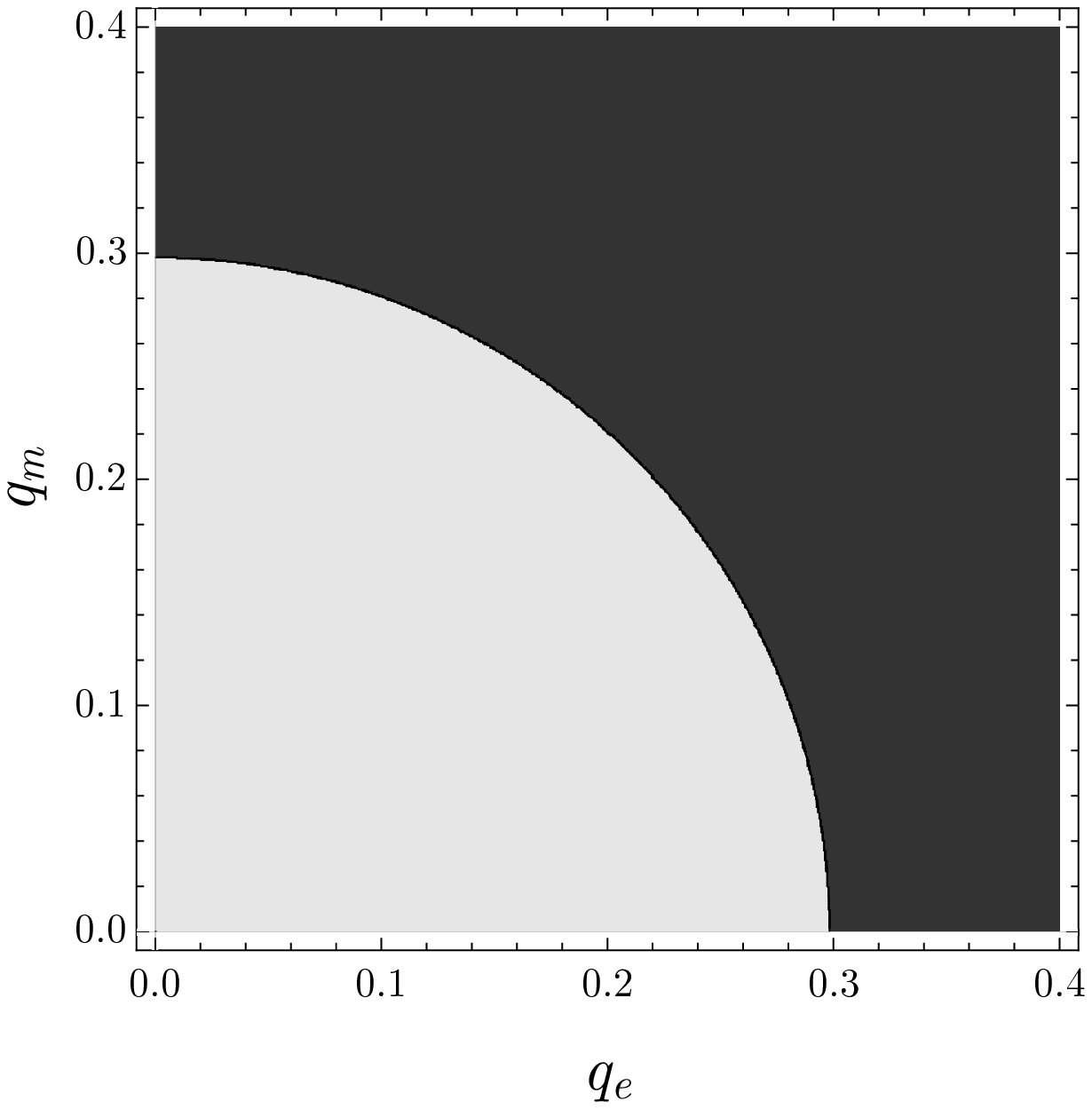} 
    \caption{$\alpha=0$} 
    \vspace{4ex}
  \end{subfigure}
  \begin{subfigure}[b]{0.5\textwidth}
    \centering
    \includegraphics[width=0.75\textwidth]{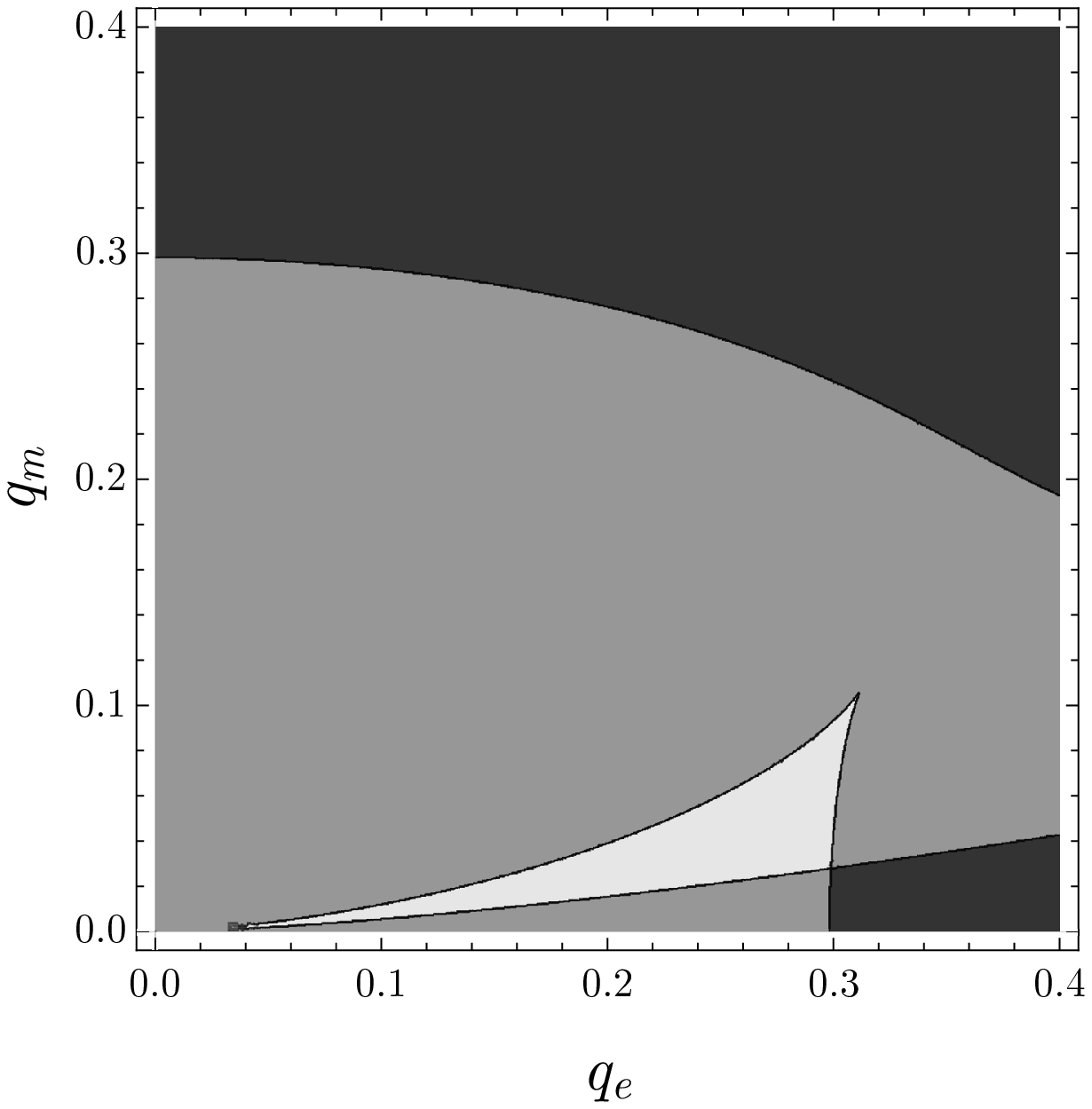} 
    \caption{$\alpha=10^{-3}$}  
    \vspace{4ex}
  \end{subfigure} 
  \begin{subfigure}[b]{0.5\textwidth}
    \centering
    \includegraphics[width=0.75\textwidth]{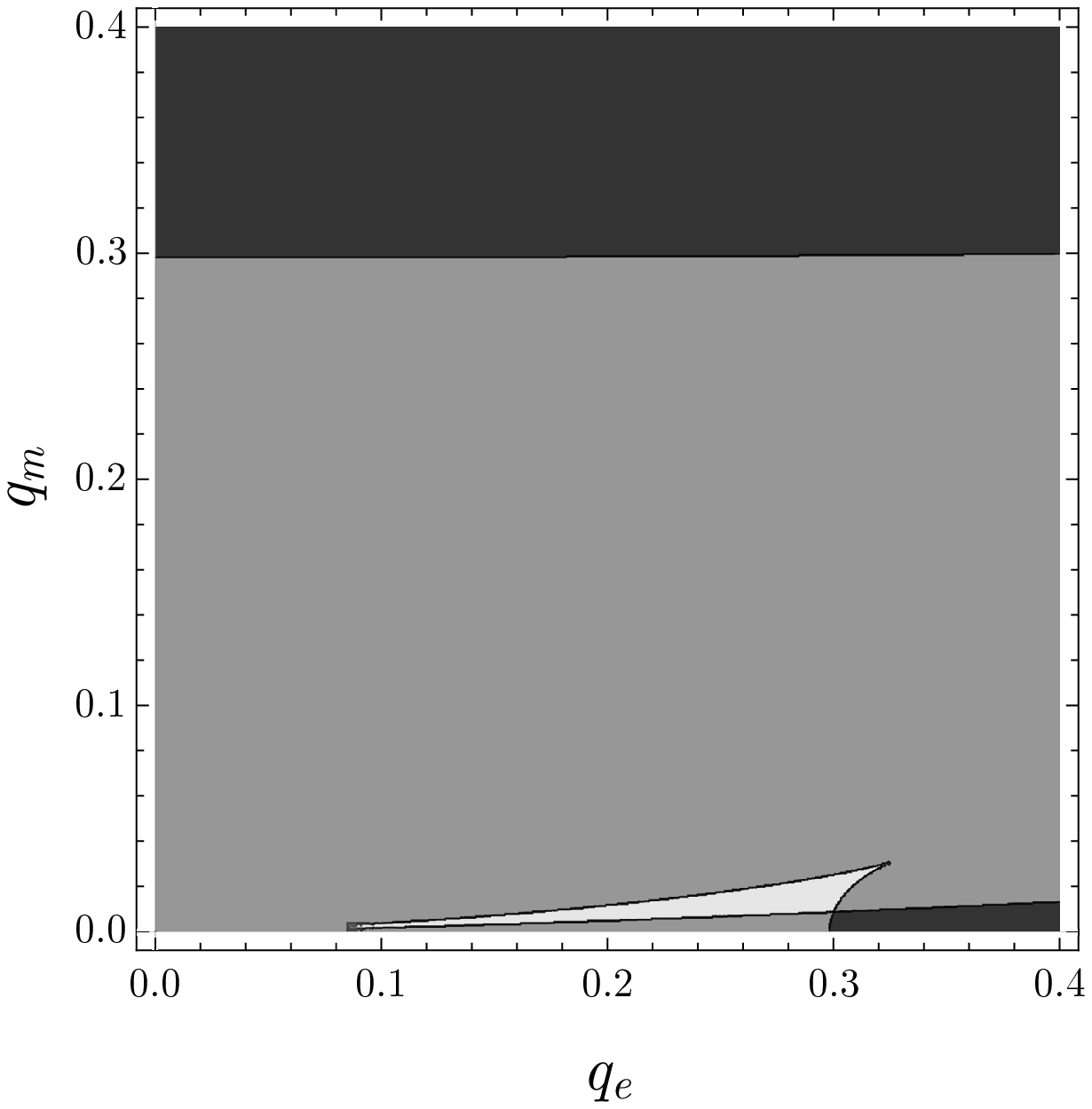} 
    \caption{$\alpha=10^{-2}$}  
  \end{subfigure}
  \begin{subfigure}[b]{0.5\textwidth}
    \centering
    \includegraphics[width=0.75\textwidth]{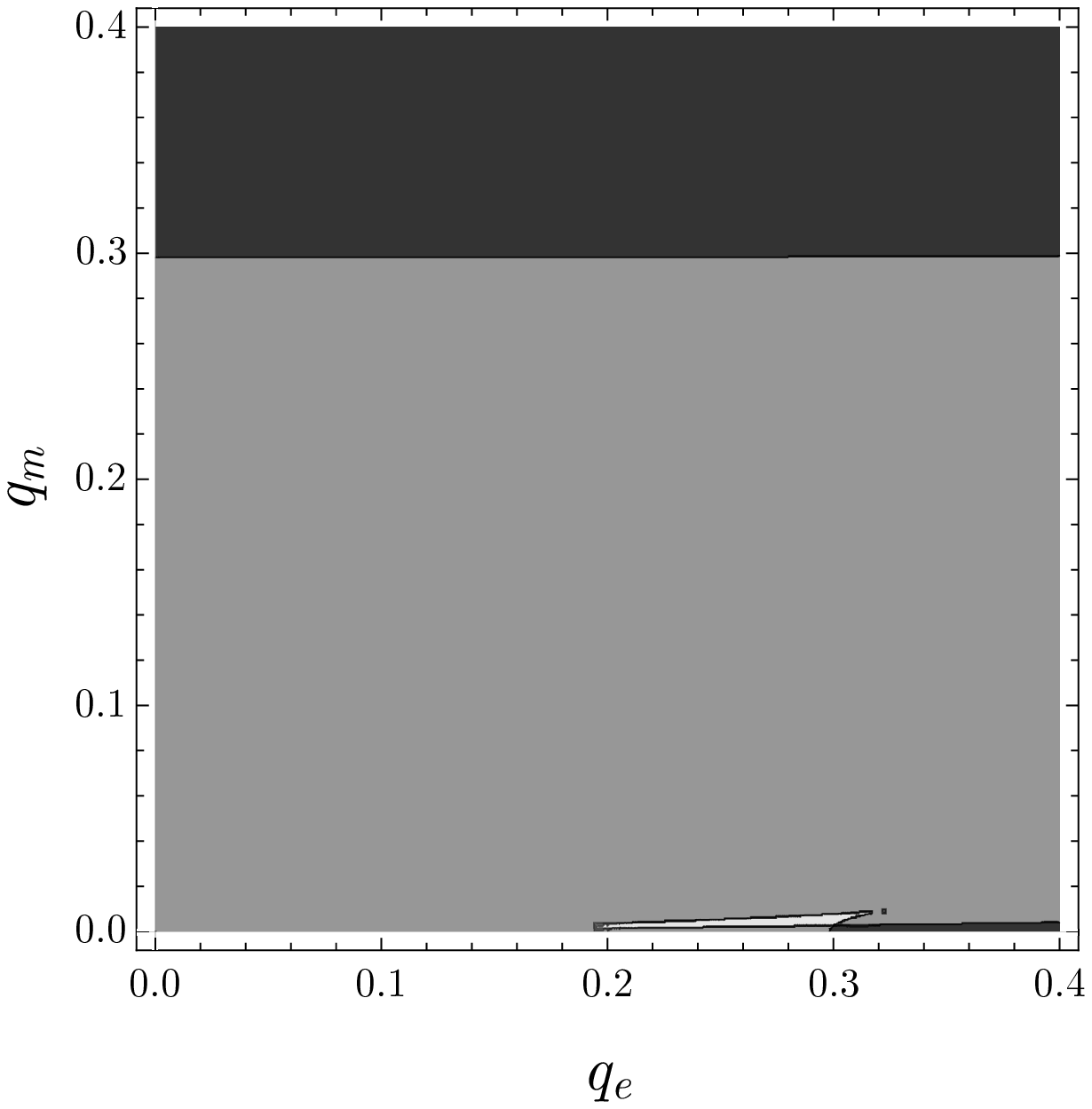} 
    \caption{$\alpha=10^{-1}$} 
  \end{subfigure} 
  \caption{Maximum number of black hole phases that may exist in various regions of the parameter space. Here, $D=5$, $\gamma=1$ and $\Lambda=-6$. Regions in black contain a single black hole, regardless of the temperature controlled by the integration constant $r_+$, which is bounded from below by the radius of the extremal black hole. For (b)--(d), regions in gray may contain at most three black holes in a given range of temperatures, while regions in white may lead to at most five configurations at a given temperature. For (a) the white region may contain at most three black holes, whereas the symmetry under the interchange $q_e\leftrightarrow q_m$ is apparent.}
  \label{fig:regions} 
\end{figure}
\newline

The entropy obeys the Bekenstein-Hawking area law. In our conventions ($16\pi G\!_N=1$), this reads
\begin{equation}
    S=\frac{A}{4G\!_N}=4\pi r_+^{D-2}\sigma_\gamma\ .
\end{equation}
The electric and the magnetic charges are given by the fluxes of the Maxwell field $F_{[2]}$ and the higher-form field strength $H_{[D-2]}$ at infinity, respectively. More precisely, $q_e$ and $q_m$ are given by
\begin{equation}
    q_e\sim\int_{\Sigma_{\infty}} \ast F_{[2]}\ ,\qquad q_m\sim\int_{\Sigma_{\infty}} H_{[D-2]}\  ,
\end{equation}
with suitable proportionality factors. With these charges, one can verify that the first principle of black hole thermodynamics is actually fulfilled; namely
\begin{equation}
    dM\, =\, T \, dS+\Phi_e\, dq_e+\Phi_m\, dq_m\ ,
\end{equation}
where the electric and magnetic potentials are
\begin{align}
    \Phi_e =& \frac{q_e\sigma_\gamma  r_+^{3-D} \, _2F_1\left[1,\frac{D-3}{2 (D-2)},\frac{7-3 D}{4-2 D},-8 \alpha q_m^2 r_+^{4-2D}   \Gamma[D-1]^2\right]}{(D-3)}\ , \\ 
    \Phi_m =&\frac{q_m^2r_+^{3-D}\sigma_\gamma}{(D-3)}+\frac{q_e^2 r_+^{3+D}\sigma_\gamma}{2(D-2)\left(r_+^{2 D}+8 \alpha   q_m^2 r_+^4 \Gamma[D-1]^2\right)q_m}\\
    &-\frac{q_e^2 r_+^{3-D}\sigma_\gamma\, _2F_1\left[1,\frac{D-3}{2 (D-2)},\frac{7-3 D}{4-2 D},-8 \alpha q_m^2 r_+^{4-2 D}   \Gamma[D-1]^2\right]}{2(D-2)q_m}\ ,
\end{align}
respectively. 

After having set the Planck length to a given value [i.e., $L_{\text{P}}=(16\pi )^{-1/(D-2)}$], there are three relevant scales to take a look at in order to study the different possible qualitative thermodynamical behaviors of the solution. These three length scales  are $L_1 = |q_m|^{2/({D-4})}$, $L_2 = |\alpha |^{1/D}$, and $\ L_3 = |\Lambda |^{-1/2}$. As is well--known, in Einstein theory in AdS, for a given temperature above certain threshold, there exist two black hole solutions, a small black hole and a large black hole, and there is a minimum temperature below which no black hole exists. This minimum temperature is fixed by the AdS curvature. Here, the Maxwell field as well as the nonlinear electromagnetic coupling modifies this picture: Fig. \ref{fig:regions} shows the maximum number of black hole phases in five dimensions for given values of the charges and the coupling $\alpha$. The details are described in the caption. In particular, it shows how the symmetry under the exchange $q_e\leftrightarrow q_m$, due to the electromagnetic duality of the $\alpha=0$ theory, gets modified as $\alpha $ increases. 
\begin{figure}[ht] 
  \begin{subfigure}[b]{0.5\textwidth}
    \centering
    \includegraphics[width=0.75\textwidth]{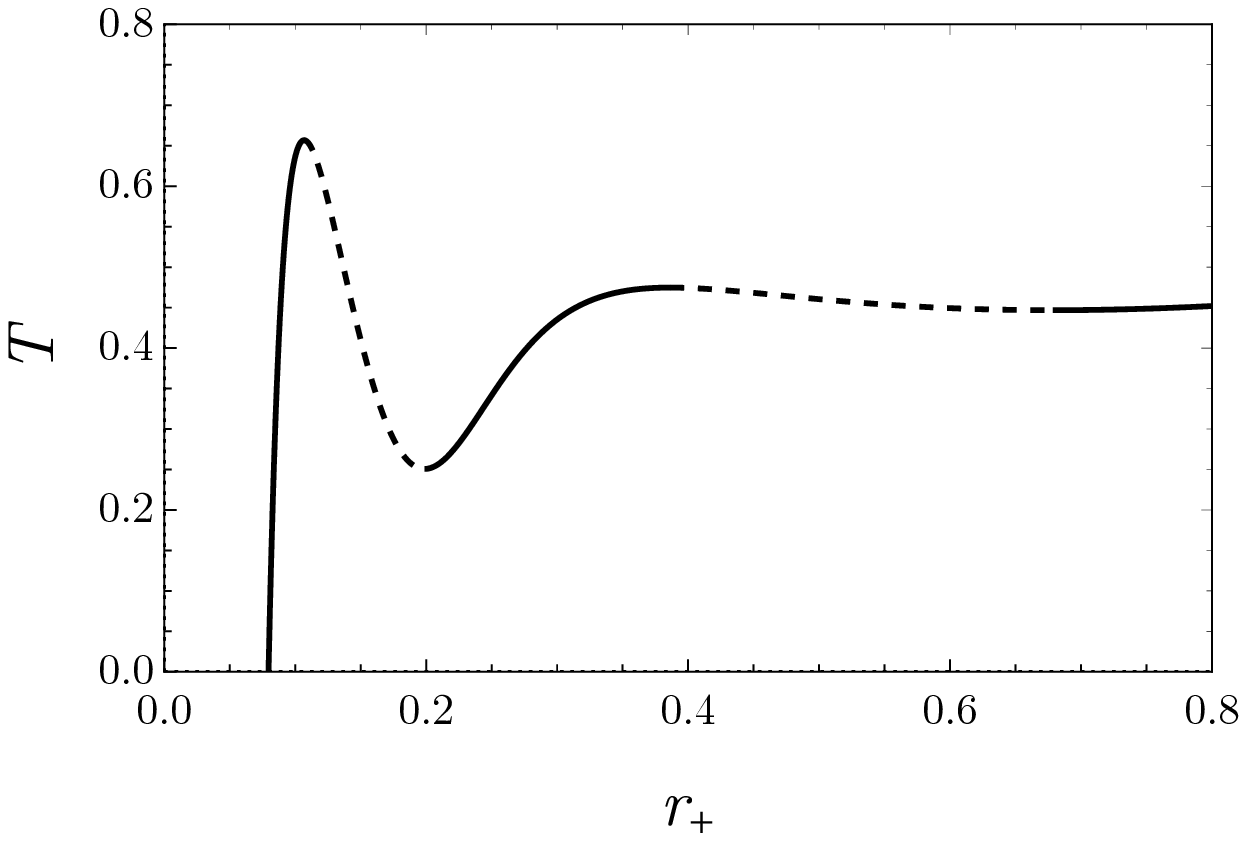} 
    \caption{} 
  \end{subfigure}
  \begin{subfigure}[b]{0.5\textwidth}
    \centering
    \includegraphics[width=0.75\textwidth]{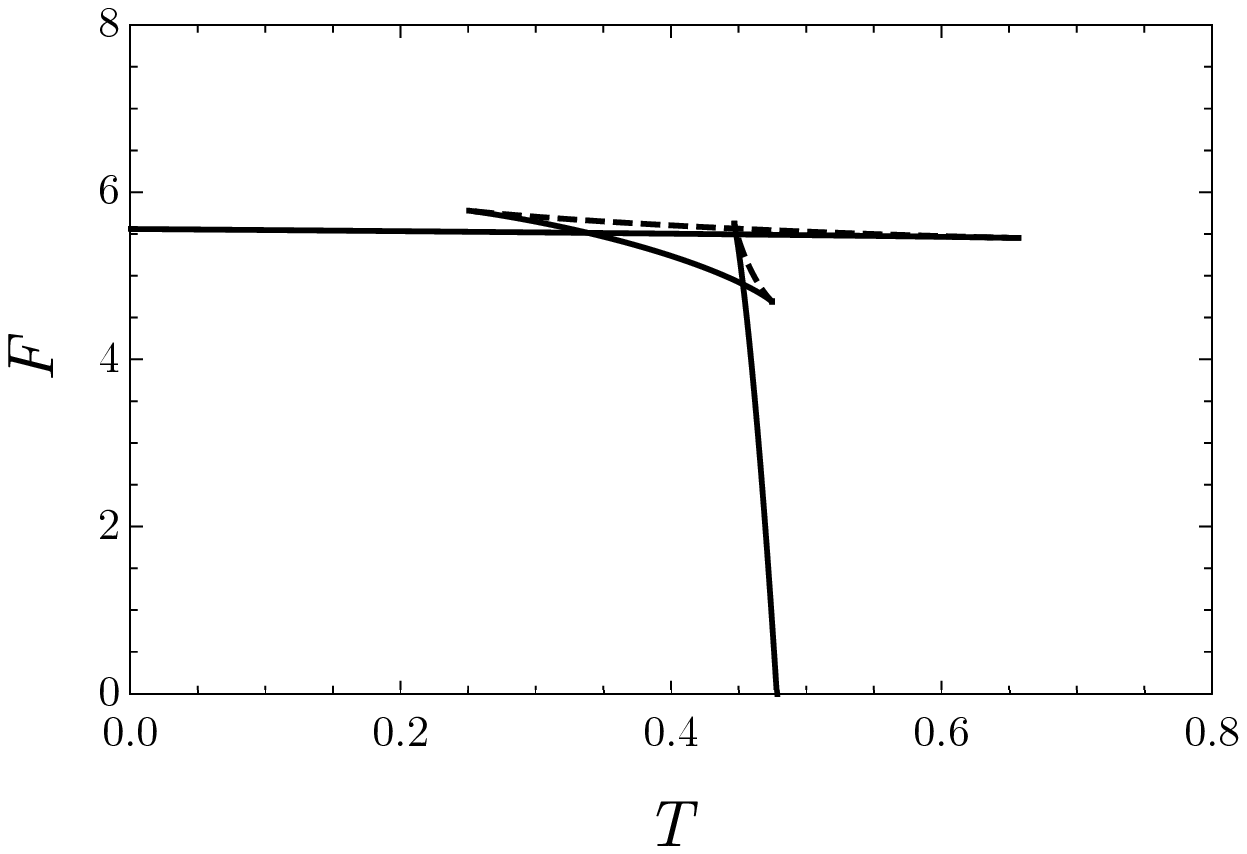} 
    \caption{}  
  \end{subfigure} 
  \caption{(a) Temperature $T$ as a function of the horizon radius $r_+$, and (b) free energy $F$ as a function of $T$ for $\alpha=10^{-3}$, $q_e=0.2$, $q_m=0.02$, $D=5$, $\gamma=1$, and $\Lambda=-6$. The solid (dashed) line corresponds to black holes with positive (negative) heat capacity.}\label{fig:FvsT}
\end{figure}
\newline

The expression for the temperature \eqref{temperature} shows that if one decreases the horizon radius, eventually the presence of the term $-q_m^2/r^{2(D-2)}$ leads to an extremal black hole, for which $T$ vanishes. Consequently, the curve in Fig. \ref{fig:FvsT} shows that there is a phase with arbitrarily low temperature. Figure \ref{fig:FvsT} also shows other features of the phase space of black holes in the canonical ensemble; in particular, one sees there that for a given range of temperatures five different configurations exist, all of them competing for the minimization of the free energy $F=M-TS$. This results in a generalized Hawking-Page picture of first-order phase transitions.

\subsection{\boldmath Four--dimensional black holes with two $U(1)$ fields}
In $D=4$ the metric of our solution reduces to the one found in Ref. \cite{Liu:2019rib}. This is expected, as the splitting of the magnetic and electric contributions of a single Maxwell field $A_{\mu}$ would resemble the coupling between a purely electric Maxwell field and another purely magnetic vector $B_{\mu }$. In our case, the explicit action with two interacting $U(1)$ fields reads 
\begin{equation}
I_4\!\left[g_{\mu\nu},A_{\mu},B_{\mu}\right]=\int\,d^4x\,\sqrt{-g}\,(R-2\Lambda)-\int\,d^4x\,\sqrt{-g}\,\left[\dfrac{1}{4}(F^2+H^2)+4\alpha H_{\mu\nu}F_{\rho\sigma}(H^{\rho\sigma}F^{\mu\nu}-4H^{\mu\rho}F^{\nu\sigma}+H^{\mu\nu}F^{\rho\sigma})\right]\label{eq:action4}\ ,
\end{equation}
leading to the field equations
\begin{eqnarray}
\mathcal{G}_{\mu\nu}&=&R_{\mu\nu}-\frac 12 R\,g_{\mu\nu}+g_{\mu\nu}\Lambda+\dfrac{1}{8}g_{\mu\nu}(F^2+H^2)-\dfrac{1}{2}(F_{\mu\rho}F_\nu{}^\rho+H_{\mu\rho}H_\nu{}^\rho)-\dfrac{\alpha}{2}g_{\mu\nu}\mathcal{L}_{\mathrm{int}}=0\ ,\\
\mathcal{M}^{\mu}&=&\nabla_{\nu}F^{\nu\mu}-16\alpha H_{\nu\rho}\nabla_{\sigma}(F^{\nu\rho}H^{\mu\sigma}-2F^{\nu\sigma}H^{\mu\rho}+2F^{\mu\nu}H^{\rho\sigma}+F^{\mu\sigma}H^{\nu\rho})\label{eq:eomm}=0\ ,\\
\mathcal{K}^{\mu}&=&\nabla_{\nu}H^{\nu\mu}-16\alpha F_{\nu\rho}\nabla_{\sigma}(F^{\nu\rho}H^{\mu\sigma}-2F^{\nu\sigma}H^{\mu\rho}+2F^{\mu\nu}H^{\rho\sigma}+F^{\mu\sigma}H^{\nu\rho})=0\ ,
\end{eqnarray}
with
\begin{equation}
\mathcal{L}_{\mathrm{int}}=4H_{\mu\nu}F_{\rho\sigma}(H^{\rho\sigma}F^{\mu\nu}-4H^{\mu\rho}F^{\nu\sigma}+H^{\mu\nu}F^{\rho\sigma})\ .
\end{equation}

Substituting the \emph{Ans\"atze} (\ref{eq:metric})--(\ref{eq:Ftens}) into these equations, we find that Eq. (\ref{eq:eomm}) is solved by
\begin{equation}
a'(r)=\dfrac{q_e r^2}{r^4+32\alpha q_m^2}\ ,  \label{potentialD}
\end{equation}
while $\mathcal{K}^{\mu}=0$ is trivially satisfied. Finally, $\mathcal{G}_{\mu\nu}=0$ is solved by the metric function
\begin{equation}\label{EinsteinsGfunction4d}
G(r)=\gamma-\dfrac{M}{2\sigma_\gamma r}-\dfrac{\Lambda r^2}{3}+\dfrac{q_m^2+ q_e^2\ {}_2F_1\bqty{1,\frac{1}{4},\frac{5}{4},-\frac{32\alpha q_m^2}{r^{4}}}}{4r^{2}}\ ,
\end{equation}
which agrees with Eq. (\ref{EinsteinsGfunction2}) for $D=4$. In fact, this coincides with the solution found in \cite{Liu:2019rib}. However, let us emphasize that, although for $D=4$ our black hole solution (\ref{EinsteinsGfunction2}) coincides with the one of \cite{Liu:2019rib}, it will generically differ from the latter in $D\neq 4$ dimensions, where the $B_{[p-1]}$ field is of a higher rank. In particular, solution (\ref{EinsteinsGfunction2}) also exists when $D$ is odd. 

Solution (\ref{EinsteinsGfunction4d}) contains the following features: The electric field (\ref{potentialD}) exhibits the screening effect which is typical of nonlinear electrodynamics. The result, however, differs from other nonlinear models such as Born-Infeld, in that the electric field vanishes at $r=0$, cf. \cite{Ferraro}. Provided $\alpha >0$, the electric field is free of singularities. At large $r$, the electric field takes the Coulombian form $\sim q_e/r^2$, as expected. The metric, on the other hand, tends to the (A)dS--Reissner--Nordstr\"{o}m geometry, where $G(r)= -\Lambda r^2/3 +\gamma -2M / r +(q_e^2+q_m^2)/(4r^2)+\mathcal{O} (1/r^3)$. We will see below that including two $U(1)$ fields is crucial to the construction of regular black holes in a suitable strongly coupled regime.

\subsection{\boldmath Causal structure\label{sec:causal}}

Let us now investigate the causal structure of the four--dimensional solution. Since the solution with $D=4$ coincides with the one found in \cite{Liu:2019rib}, one could simply refer to that reference for analysis of the horizon structure. Nevertheless, let us present here a detailed, different analysis to determine the horizon structure analytically. To do so, it is convenient to define an auxiliary function 
\begin{equation}
Y(r)=2r\sigma_\gamma G(r)+M=\frac{\sigma_\gamma}{2r}\pqty{4r^2\gamma+q_m^2+{}_2F_1\bqty{\frac{1}{4},1,\frac{5}{4},-\frac{32q_m^2\alpha}{r^4}}q_e^2}\ ,
\end{equation}
and investigate its extrema. These are located at the solutions of $G+rG'=0$. For simplicity, we have set the bare cosmological constant to zero, i.e. $\Lambda=0$, so we will be dealing with asymptotically flat black holes. Also, we will restrict our attention to the case of spherical horizon, i.e. $\gamma=1$. The positive function $Y(r)$ goes as $\sim q_m^2/r+\mathcal{O}(r)$ near the origin, while it asymptotically behaves as $\sim r+\mathcal{O}(1/r)$. This means that a single extremum is necessarily a global minimum, two extrema are the value at a saddle point and a global minimum, etc. After a change of variables $z=r^2$, one ends up asking for the solutions of the cubic equation
\begin{equation}
4 z^3-(q_m^2+q_e^2)z^2+128\alpha q_m^2 z -32\alpha q_m^4=0\ .
\end{equation}
The optimal expression for this cubic equation is its depressed form which is achieved by the further change of variables $z=\tilde{z}+(q_m^2+q_e^2)/12$. This leads us to
\begin{equation}
W(\tilde{z}):=\tilde{z}^{3}+c_1\tilde{z}+c_2=0\ 
\end{equation}
with
\begin{equation}
c_1=32\alpha q_m^2-\frac{(q_m^2+q_e^2)^2}{48}\ ,\quad c_2=\frac{8\alpha q_m^2(q_e^2-2q_m^2)}{3}-\frac{(q_m^2+q_e^2)^3}{864}\ .
\end{equation}
Since $z=r^2$, the sensible roots of $W(\tilde{z})$ will be real and positive. To move on, one needs to separate cases according to the behavior of the discriminant $\Delta=-(4c_1^3+27c_2^2)$, namely (i) $\Delta=0$, (ii) $\Delta>0$ and (iii) $\Delta<0$. 

For the case (i) there exist two subcases, according to whether $c_1=0$ or not. If $c_1=0$, then $c_2=0$, and this can happen only in the particular configuration $q_e=2\sqrt{2}q_m$ and $q_m=\sqrt{512\alpha/27}$. Consequently, zero is a triple root which can be traced back to the $r$ coordinate via the chain of backward transformations
\begin{equation}
\tilde{z}_{\star}=0\to z_\star=\frac{128\alpha}{9}\to r_\star=\sqrt{\frac{128\alpha}{9}}\ .
\end{equation}
A minimum, $M_\star\equiv Y(r_\star)\sim 100\pi\sqrt{\alpha}$, becomes the necessary mass bound for the formation of a black hole, as for $M<M_\star$ the singularity at $r=0$ is naked. When the inequality is saturated, an extremal black hole forms with its horizon located at $r_\star$, while for $M>M_\star$ there exist two horizons. Now, if $c_1\neq 0$, one finds a single root $r_1$ and a double root $r_\star$, 
\begin{equation}
r_1=\frac{1}{2}\sqrt{\frac{(q_m^2+q_e^2)+512 q_m^2(5q_m^2-4 q_e^{2})}{\mathcal{C}}}\ ,\quad r_\star=8\sqrt{\frac{\alpha q_m^2(q_e^2-8q_m^2)}{\mathcal{C}}}\ ,\quad \mathcal{C}:=(q_m^2+q_e^{2})^2-1536\alpha q_m^2\ ,
\end{equation}
both sensible in the parameter domain where the reality of the square root is guaranteed. The parameters are also subject to the constraint
\begin{equation}
(64q_m)^2\alpha=q_e^2(q_e^2+20q_m^2)-8q_m^4\pm \sqrt{q_e^2(q_e^2-8q_m^2)^3}\ ,
\end{equation} 
coming from the vanishing of the discriminant. In this case, $r_\star$ is the saddle point (associated with a mass $M_\star\equiv Y(r_\star)$) which is strictly greater than $r_1$, the location of the minimum $M_1\equiv Y(r_1)$. Again, the minimum represents the smallest mass necessary for the formation of a black hole, while when $M_1<M<M_\star$ or $M>M_\star$ there exist two horizons. When $M=M_1$, the two horizons coalesce at $r_1$, while when $M=M_\star$ the outer horizon is located at $r_\star$. Additionally, $M_1\neq M_\star$ always. 

When $\Delta>0$, case (ii), $W$ has three positive real roots which in terms of the $r$ coordinate are expressed as
\begin{equation}
r_k=\Bqty{\frac{q_m^2+q_e^2}{12}-\frac{1}{6}\sqrt{\mathcal{C}}\cos(\frac{1}{6}\bqty{\pi(4k+1)+2\asin(\frac{(q_m^2+q_e^2)^3-2304\alpha q_m^2(q_e^2-2q_m^2)}{\mathcal{C}^{3/2}})})}^{1/2}\ ,
\end{equation}
for $k=1,2,3$. In the suitable region of the parameter space, it holds that $r_1>r_2>r_3$. Consequently, $M_1\equiv Y(r_1)$ and $M_3\equiv Y(r_3)$ are local minima, while $M_2\equiv Y(r_2)$ is a global maximum. Let us use the notation $M_{\mathrm{min}}=\min(M_1,M_3)$ and $M_{\mathrm{max}}=\max(M_1,M_3)$. Again, when $M<M_{\mathrm{min}}$ the singularity is naked, while for $M_{\mathrm{min}}<M<M_{\mathrm{max}}$ there exist two horizons. In the region $M_{\mathrm{max}}<M<M_2$ we find a total of four horizons, while for $M>M_2$ the number reduces to two. When $M=M_{\mathrm{min}}$, an extremal horizon forms at $r_\mathrm{min}$, whereas when $M=M_{\mathrm{max}}$ we have three horizons, the innermost at $r_{\mathrm{max}}$ being extremal. Three horizons exist also when $M=M_2$ where now the intermediate one is formed at $r_2$. Of special interest is also the case when $M_{\mathrm{min}}=M_{\mathrm{max}}\equiv M_{\star}$. The smallest black hole is of mass $M_\star$, and it possesses two extremal horizons located at $r_1$ and $r_3$. Then, for masses $M>M_\star$ the behavior follows the unsaturated case.

Finally, for the case (iii), $W$ has only one single positive real root, the manifest expression of which depends on the sign of $c_1$. As an example, we give the root when $c_1<0$:
\begin{equation}
r_1=\Bqty{\frac{q_m^2+q_e^2}{12}+\frac{1}{6}\sqrt{\mathcal{C}}\cosh(\frac{1}{3}\mathrm{acosh}\pqty{\frac{(q_m^2+q_e^2)^3-2304\alpha q_m^2(q_e^2-2q_m^2)}{\mathcal{C}^{3/2}}})}^{1/2}\ .
\end{equation}
Here, the smallest black hole is an extremal one with mass $M_1\equiv Y(r_1)$ and its horizon formed at $r_1$. Then, for $M>M_1$ we have the phase of two horizons. With regard to constant curvature asymptotics, there is no qualitative difference since the maximum number of positive real roots of the quartic equation $Y'=0$ is still three, and the solution qualitatively exhibits the same behavior as for the cubic $Y$.

\subsection{\boldmath Three--dimensional black holes\label{sec:four}}
In $D=3$ dimensions, the Lagrangians introduced above reduce to Einstein-Maxwell theory with a cosmological constant plus a scalar field $\chi$. The scalar field has a nonminimal coupling with the $U(1)$ field and interacts with a purely electric stress tensor $F_{\mu\nu}$ via a term of the form
\begin{equation}
\delta^{\alpha_1\alpha_2\alpha_3}_{\beta_1\beta_2\beta_3}F_{\alpha_1\alpha_2}\nabla_{\alpha_3}\chi F^{\beta_1\beta_2}\nabla^{\beta_3}\chi\ ,
\end{equation} 
so that the complete action takes the form 
\begin{equation}
I_3\bqty{g_{\mu\nu},A_\mu,\chi}=\int\,d^3x\,\sqrt{-g}\,(R-2\Lambda)-\int\,d^3x\,\sqrt{-g}\,\left[\dfrac{1}{4}F^2+\dfrac{1}{2}(\nabla \chi)^2+\alpha\mathcal{L}_{\mathrm{int}}\right]\ .\label{eq:3dModel}
\end{equation}
The interaction term explicitly reads
\begin{equation}
\mathcal{L}_{\mathrm{int}}=2F^2(\nabla\chi)^2-4F^{\mu}{}_{\rho}F^{\nu\rho}\nabla_{\mu}\chi \nabla_{\nu}\chi \ , \label{int3d}
\end{equation}
and the field equations are
\begin{eqnarray}
\mathcal{G}_{\mu\nu}&=&G_{\mu\nu}+g_{\mu\nu}\Lambda -\frac{1}{2}F_{\mu\rho}F_\nu{}^\rho+\dfrac{1}{8}g_{\mu\nu}F^2-\frac{1}{2}\nabla_\mu\chi \nabla_\nu \chi+\frac{1}{4}g_{\mu\nu}(\nabla\chi)^2-\dfrac{\alpha}{2}g_{\mu\nu}\mathcal{L}_{\mathrm{int}}=0\ ,\\
\mathcal{M}^{\mu}&=&\nabla_{\nu}F^{\nu\mu}+24\alpha \nabla_\nu\chi \nabla_\rho(F^{[\mu\nu}\nabla^{\rho]}\chi)=0\ ,\label{eq:maxwell3}\\
\mathcal K&=&\Box\chi+4\alpha F_{\mu\nu}\nabla_\rho(F^{\mu\nu}\nabla^{\rho}\chi-2F^{\mu\rho}\nabla^{\nu}\chi)=0\ .
\end{eqnarray}

The spacetime metric we consider is of the form (\ref{eq:metric}) with $D=3$ and, in such a case, $\gamma=0$. Again, $F_{\mu\nu}$ is purely electric, like in Eq. (\ref{eq:Ftens}), while we assume the simplest linear \emph{Ansatz} for the scalar, i.e., $\chi(x)=\beta x$ for an arbitrary real parameter $\beta$; the Klein--Gordon equation is then identically solved. Notice that $d\chi\sim dx$ which implies that the exterior derivative of the scalar is proportional to the volume form of the $t,r$--constant manifold. Having said that, we can integrate Eq. (\ref{eq:maxwell3}) to find the electric field
\begin{equation}
a'(r)=\dfrac{q_e r}{r^2+8\alpha \beta^2}\ ,
\end{equation}
where we observe that the constant parameter $\beta$ effectively plays the role of the magnetic charge in the previous examples. Finally, substituting all results back into $\mathcal{G}_{\mu\nu}$, we can solve the metric field equations, obtaining the solution 
\begin{equation}
G(r)=-\frac{M_0}{2\pi }-\Lambda r^2- \dfrac{q_e^2}{4}\log (r^2+8\alpha\beta^2)  -\dfrac{\beta^2}{2}\log r\ ,\label{Ultima}
\end{equation}
which deforms the electrically charged BTZ solution \cite{Banados:1992wn} with noncompact horizon $\Sigma_1=\mathbb{R}$ and $d\Sigma_1^2=dx^2$. Notice also that in the absence of the interaction term ($\alpha=0$), both the scalar and the Maxwell field contribute in the same manner to the lapse function, since they can be mapped by Hodge duality in this case.

\subsection{Nonsingular solutions}

We have extended the quasitopological electromagnetic Lagrangians introduced in \cite{Liu:2019rib} by adding higher-rank fundamental forms $B_{[p-1]}$. This field, being independent of the Maxwell field $A_{\mu}$, allows us to construct a family of nonsingular black hole solutions, even in four dimensions. Originally, regular black holes were geometrically constructed in \cite{Bardeen}, and the embedding of such black holes in a dynamical theory was successfully achieved in \cite{AyonBeato:1998ub,AyonBeato:1999ec,AyonBeato:1999rg}; for recent realizations see \cite{Babichev:2020qpr}, and for a review see \cite{Frolov:2016pav} and references therein.

We will then demand the spherically symmetric metric to approach a constant curvature background near the origin, which as seen below can be achieved in a suitable strongly coupled regime. In the region $r\rightarrow 0$, we require
\begin{equation}
G(r)= 1-\frac{r^2}{l_{\mathrm{eff}}^2}+\mathcal{O}(r^3)\ ,
\end{equation}
which suffices to guarantee a regular behavior at the origin, so that the Riemannian curvature remains finite there. This also ensures the completeness in the geodesic sense \cite{sakharov}. Here, we will see that a family of such spacetimes is possible in the setup discussed above. 

Let us begin by studying the stress tensor of the theory, $T^{a}_{\, b}=\mathrm{diag}(-\rho,-\rho,p_{x_1},\ldots, p_{x_p})$, where $\rho$ can be read off from Eq. (\ref{eq:density}) given the component of the energy-momentum tensor projected on a locally orthonormal basis, i.e. $T^{a}_{\, b}=e^{a}_\mu e_b^{\nu} T^{\mu}_\nu$, with $g_{\mu\nu}=e^a_{\ \mu}e^b_{\ \mu}\eta_{ab}$. As said before, $\rho>0$ provided $\alpha >0$, this being a requirement for a regular electric field, everywhere. Moreover, $\rho(r)$ is a monotonically decreasing function, falling off fast enough as to provide a finite ADM mass, as discussed above. On the other hand, the energy density of the matter fields, still diverges at the origin due to the magnetic field contribution $\sim q_m^2/(4 r^{2(D-2)})$ [see Eq. \eqref{eq:density}], which comes from the kinetic term $H^{2}$. Considering a strongly coupled regime one can disregard such a kinetic term, which leads to a finite energy density at the origin. This limit can be formally taken in the solution by sending $\alpha$ to infinity while keeping $\sqrt{\alpha}q_m$ finite. The metric function therefore reads
\begin{equation}\label{EinsteinsGfunction2noH2}
G(r)=1-\dfrac{M}{(D-2)\sigma_1 r^{D-3}}-\dfrac{2\Lambda r^2}{(D-1)(D-2)}+\dfrac{q_e^2\ {}_2F_1\bqty{1,\frac{D-3}{2(D-2)},\frac{3D-7}{2(D-2)},-\frac{8\alpha q_m^2\Gamma[D-1]^2}{r^{2(D-2)}}}}{2(D-2)(D-3)r^{2(D-3)}}\ .
\end{equation}
Expanding at short distances, one finds
\begin{equation}
G(r)= 1 + \frac{\mathcal{E}-M}{r^{D-3}}-\frac{r^2}{l_{\mathrm{eff}}^2}+\mathcal{O}\left(r^3\right)\ ,
\end{equation}
where we have defined
\begin{equation}
{l_{\mathrm{eff}}^{2}}:={16(D-1)(D-2)} \, \pqty{32\Lambda+\frac{q_e^2}{\alpha q_m^2 \Gamma[D-1]^2}}^{-1}\ ,
\label{lefforigin}
\end{equation}
and
\begin{equation}
\mathcal{E}=q_e^2\frac{2^{\frac{17-7D}{2(D-2)}}\sigma_1\Gamma\bqty{\frac{D-3}{2(D-2)}}\Gamma\bqty{\frac{D-1}{2(D-2)}}}{(D-2)(\alpha q_m^2 \Gamma[D-1]^2)^{\frac{D-3}{2(D-2)}}}\ .
\end{equation}

Therefore, for the solution to be regular, we need to fix the ADM mass in terms of a combination of the electric and magnetic charges, namely $M=\mathcal{E}$, leading also to a relation between the mass and the nonvanishing energy density of the matter fields at the origin \cite{Spallucci:2017aod}. 
\begin{figure}[ht!]
\centering
\includegraphics[scale=.75]{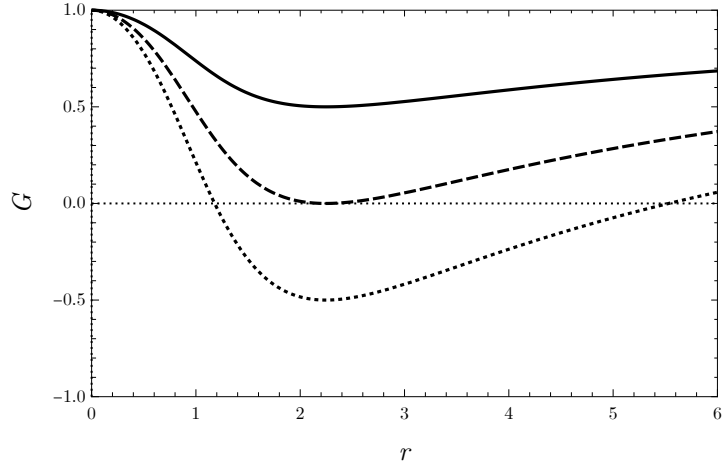}
\caption{Showing cases with flat asymptotics: gravitational soliton with a regular origin, nonsingular extremal black hole, and nonsingular black hole with two horizons.}
\label{fig:regBH}
\end{figure}
\newline
One can check that all the components of the Riemann tensor $R^{\mu\nu}{}_{\rho\sigma}$ are finite at the origin, and therefore each algebraic curvature invariant of order $k$ takes a value $\sim l_{\mathrm{eff}}^{-2k}$ at $r=0$. The possible horizon structures of these singularity-free black holes can be read from Fig. \ref{fig:regBH}. The absence of an event horizon leads to a gravitational soliton with a regular origin. Finally, as a closing remark, it would be interesting to apply our approach to theories like~\cite{Feng:2015sbw,Cisterna:2020kde} where a nonminimal coupling to gravity is present.

\section*{Acknowledgments} 
The work of A. C. is supported by Fondecyt Grant No. 11170274 and Proyecto Interno Ucen I+D-2018 CIP 2018020. K. P. acknowledges financial support provided by the European Regional Development Fund through the Center of Excellence TK133 ``The dark side of the Universe" and PRG356 ``Gauge gravity: Unification, extensions and phenomenology." The work of G. G. is supported by CONICET through Grant No. PIP 1109 (2017). J. O. is supported by FONDECYT Grant No. 1181047.

\end{document}